\documentclass[12pt]{iopart}
\usepackage{graphicx}
\usepackage{bm}
\begin{document}
\title[Extreme value theory and the St. Petersburg paradox in the failure statistics of wires]{Extreme value theory and the St. Petersburg paradox in the failure statistics of wires}

\author{Alessandro Taloni$^1$, Stefano Zapperi$^{2,3}$}

\address{$^1$ CNR-ISC, Institute for Complex Systems, Via dei Taurini 19, 00185 Roma, Italy}
\address{$^2$ Center for Complexity and Biosystems and Department of Physics, University of Milano, Via Celoria 16, 20133 Milano, Italy}
\address{$^3$ CNR-ICMATE, Institute of Condensed Matter Chemistry and Technologies for Energies, Via Roberto Cozzi 53, 20125 Milano, Italy}
\ead{alessandro.talonir@isc.cnr.it, stefano.zapperi@unimi.it}

\begin{abstract}
The fracture stress of materials typically depends on the sample size
and is traditionally explained in terms of extreme value statistics. A recent work reported results on the carrying capacity of long polyamide and polyester wires and interpret the results in terms of a probabilistic argument known as the St. Petersburg paradox. Here, we show that the same results can be better explained in terms of extreme value statistics. We also discuss the relevance of rate dependent effects. 
\end{abstract}
\submitto{\jpg}
\maketitle
\normalsize

\section*{Introduction}
The failure of macroscopic objects is determined both by intrinsic properties, such as its crystalline structure, and by extrinsic factors as the grain boundaries, vacancies, dislocations or microcracks. In particular, the strength of brittle materials is largely influenced by the presence of flaws, as it was first demonstrated by Griffith in his seminal paper 100 years ago \cite{griffith1921vi}. The Griffith's criterion is the foundation stone of fracture mechanics, showing, by simple linear elasticity arguments, that a material fails at a lower nominal applied stress when a crack is present. 

The Griffith theory is also relevant to describe fracture size effects or the observation that larger samples are generally weaker. In fact, the larger the sample, the higher is the probability to find microcraks
that might become unstable when a threshold stress is reached. According to the Griffith's theory this threshold is inversely proportional to the square root of defect size. Thus the strength of a material corresponds to the stress needed to trigger the first crack instability event, namely the smallest threshold corresponding to the largest flaw. In this sequence of arguments lies the connection between fracture mechanics , materials failure and extreme value statistics 
(for a review see \cite{taloni2018size}).

In a recent letter \cite{ball2020classic,fontana2020petersburg}, the authors proposed a connection between materials failure statistics and the St. Petersburg paradox by linking the average load carrying capacity of a wire to its length. The result, however, was derived assuming that "the force required to fracture the fiber is a linear function of the defect size" \cite{fontana2020petersburg}, which is in glaring contrast with fracture mechanics. Here we address the problem combining extreme value theory (EVT) \cite{taloni2018size} with the Griffith's stability crack criterion \cite{griffith1921vi}.  We also show that in the asymptotic limit, the wire strength  follows the Gumbel's distribution, in full agreement with the data reported
in \cite{fontana2020petersburg}, as we demonstrate using the maximum likelihood method. We thus conclude that the load carrying capacity of the wires studied in \cite{fontana2020petersburg} follows EVT, in agreement with previous observations for different materials \cite{taloni2018size}.

\section{Theory}

Let us consider a wire of length $L$ which we divide in $N=L/L_0$ independent elements of size $L_0$. Following Ref. \cite{fontana2020petersburg}, we want to relate the statistics of the micro-cracks present in the wire with its failure strength.

Defining $P(n)$ as the probability density function (pdf) of micro-cracks of length $w \equiv n L_0$,  we introduce

\begin{equation}
F(z)=\int_0^z dn P(n)
\label{cumulative_crack}
\end{equation}

\noindent as the probability that no micro-crack larger than z$L_0$ will be found in the wire. 
Next, we look for the statistics of  the largest micro-crack in the wire  $w_{nmax}=n_{max}L_0$ which satisfies  the only constraint  $w_{max}=n_{max}L_0\ll L$ \cite{duxbury1987breakdown}. If $\rho_N(n_{max})$ is the pdf
for the largest micro-crack, then 

\begin{equation}
F_N(z)=\int_0^z dn_{max}\, \rho_N(n_{max})=\left[F(z)\right]^N
\label{cumulative_crack_N}.
\end{equation}

 \noindent The Fisher-Gnedenko-Tippet theorem ensures that $F_N(z)\to G(z)$ for large $N$, where $G$ belongs to one of three families only: Weibull, Fr\'echet or Gumbel \cite{fisher1928limiting, gnedenko1943distribution}. The convergence to either one of these universal distributions depends on the asymptotic properties of $P(n)$ \cite{leadbetter2012extremes,manzato2012fracture}. If the distribution of micro-cracks has an exponential tail \cite{kunz1978essential,manzato2012fracture,duxbury1987breakdown}, $G(z)$ converges asymptotically to the Gumbel distribution \cite{gumbel2004statistics}: 
 
\begin{equation}
F_N(z)\to G(z)=e^{-Ne^{-\frac{z-\mu}{\beta}}}
\label{gumbel_cumulative}
\end{equation} 
 
  To derive the
fracture strength, the authors of Ref. \cite{fontana2020petersburg} assume that it is linearly dependent on the size of the largest defect ($n_{max}$), obtained through the analogy with the 
St. Petersburg model. However, this assumption is not justified by fracture mechanics. A relation between crack length and fracture strength in an elastic medium is provided by the Griffith's stability criterion. It states that a crack of length $w$ subject to a normal stress $\sigma$ is stable as long as $\sigma<K_{1C}/Y\,w^{-1/2}$\cite{griffith1921vi}, where $K_{1C}$ is the
critical stress intensity factor and $Y$ is a geometric factor. In our context, the wire should break when the largest micro-crack becomes unstable, i.e. when it is reached the stress

\begin{equation}
\sigma_{break}=K_{1C}/Y\,w_{max}^{-1/2}
\label{griffith_strength}.
\end{equation}

\noindent This relation corresponds to the weakest link hypothesis in elastic mechanics \cite{taloni2018size}. Thanks to \ref{griffith_strength}, we can formally write down the probability that a wire of length $L$ brakes at stress $\sigma_{break}$ as

\begin{equation}
\rho_L(\sigma_{break})=-\frac{2K_{1C}^2}{Y^2L_0\sigma_{break}^3}\rho_N\left(n_{max}=\frac{K_{1C}^2}{Y^2L_0\sigma_{break}^2}\right), 
\end{equation}

\noindent and, by means of \ref{cumulative_crack_N}, introduce the probability that the wire does not fail up to a stress $\sigma$

\begin{equation}
\Sigma_L(\sigma) = F_N(z) 
\label{survival_crack_L}
\end{equation}

\noindent with $z=\frac{K_{1C}^2}{Y^2L_0\sigma^2}$. For large samples, the convergence theorem \ref{gumbel_cumulative} yields

\begin{equation}
 \Sigma_L(\sigma)\sim e^{-\frac{L}{L_0}\,e^{-\left(\frac{\sigma_0}{\sigma}\right)^2}},
\label{DLB}   
\end{equation}

\noindent with $\sigma_0=\frac{K_{1C}}{Y\sqrt{L_0\beta}}$.  This is the Duxbury-Leath-Beale distribution \cite{duxbury1987breakdown}, which was shown to converge to the Gumbel distribution as  $L\gg L_0$ , i.e.

\begin{equation}
\Sigma_L(\sigma)\to e^{-\frac{L}{L_0}\,e^{\frac{\sigma-\mu}{\beta}}}
\label{Gumbel_failure_surv_L}
\end{equation}

\noindent \cite{manzato2012fracture}. Hence, in analogy with the expression \ref{cumulative_crack_N}, we can define the survival probability for a single crack as

\begin{equation}
\Sigma_0(\sigma)=e^{-e^{\frac{\sigma-\mu}{\beta}}}
\label{Gumbel_failure_surv}.
\end{equation}

\noindent Finally, the  average breaking stress of a wire of length $L$ is  given by

\begin{equation}
    \langle \sigma\rangle = \int_0^{\infty}d\sigma \Sigma_L(\sigma)= \beta\left[\gamma -\ln\left(L/L_0\right)\right]+\mu,
\end{equation}

\noindent which recovers Eq.(1) of \cite{fontana2020petersburg} and
 $\gamma$ is the Euler-Mascheroni constant.

\section{Comparison with the experiments}

In Ref.\cite{fontana2020petersburg} tensile experiments were performed on  polyester and polyamide wires. In a first series of experiments the fracture strengths of samples ranging from 1 mm to 1 km were reported. A second sequence of experiments showed a small strain rate dependence  of the fracture forces for wires of fixed lengths, in both materials. 

\noindent To capture and reinterpret the experimental results within the framework of the classical fracture mechanics, we have used the Gumbel form of the generalized EVT, which is purposely designed to account for strain and thermal effects  \cite{taloni2018size,sellerio2015fracture}. In this context we can introduce the survival distribution function as the probability that a wire of volume $V_i$ is still intact up to a stress $\sigma_i$, if the tensile experiment is conducted at a strain rate $\dot{\varepsilon_i}$ and temperature $T$,

\begin{equation}
 \Sigma_{V_i}(\sigma_i;\dot{\varepsilon_i},T)=\left[\int_{\sigma_i/E}^{\infty} d\varepsilon_f \rho_0(\varepsilon_f) S_0(\sigma_i;\dot{\varepsilon_i},T|\varepsilon_f)\right]^{\frac{V_i}{V_0}}.
\label{Sigma_V_thermal}   
\end{equation}

\noindent The quantity $\rho_0(\varepsilon_f)$ corresponds to the Gumbel failure strain probability density function

\begin{equation}
\rho_0(\varepsilon_f)=\frac{e^{\frac{E\varepsilon_f-\mu}{E\varepsilon_0}-e^{\frac{E\varepsilon_f-\mu}{E\varepsilon_0}}}}{\varepsilon_0},
\label{Gumbel_strain_pdf}
\end{equation}

\noindent obtained  by differentiating the expression \ref{Gumbel_failure_surv} ($\rho_0(\sigma)=-\frac{d\Sigma_0(\sigma)}{d\sigma}$) and substituting $\sigma=E\varepsilon_f$ and $\beta=E\varepsilon_0$, with $E$ Young's modulus.  The thermal factor $S_0$ can be derived from the Kramer's theory for the transition rate as \cite{sellerio2015fracture}

\begin{equation}
S_0(\sigma_i;\dot{\varepsilon_i},T|\varepsilon_f)=e^{-\frac{\omega_0}{\sqrt{2\pi \dot{\varepsilon_i}}}\sqrt{\frac{EV_0}{k_BT}}\left[e^{-\frac{EV_0}{2k_BT}\left(\varepsilon_f-\sigma_i/E\right)^2}-e^{-\frac{EV_0}{2k_BT}\varepsilon_f^2}\right]},
\label{thermal_factor}
\end{equation}

\noindent where $\omega_0$ is a characteristic frequency and $k_B$ the Boltzmann's constant. 

\noindent The data reported in \cite{fontana2020petersburg} correspond to the fracture forces $F_i$ for fibers of lengths $L_i$. Each experiment was performed at a different strain rate, such that $\dot{\varepsilon_i}=\frac{\alpha}{L_i}$, where $\alpha=6\  10^{-3} m/sec$ if $L_i<1 m$ or $\alpha=71\  10^{-3} m/sec$ if $L_i\geq1 m$. Under these conditions, the proper way of fitting the experimental fracture data is the Maximum Likelihood (ML) method \cite{taloni2018size}, which consists in determining the values of the parameters $\mu,\varepsilon_0,V_0,\omega_0$ that maximize the function

\begin{equation}
\ln \mathcal{L} = \sum_{i=1}^N\ln \rho_{V_i}(\sigma_i;\dot{\varepsilon_i};T).
\label{ML}
\end{equation}

\noindent $\rho_{V_i}(\sigma_i;\dot{\varepsilon_i};T)=-\frac{d\Sigma_{V_i}(\sigma_i;\dot{\varepsilon_i};T)}{d\sigma_i}$ is the probability that a wire of volume $V_i$ fails at stress $\sigma_i$, under the specific experimental conditions $(\dot{\varepsilon_i};T)$. Each volume was calculated as $V_i=\pi L_i\left(\frac{d}{2}\right)^2$ with $d=0.25\ 10^{-3} m$ for polyester fibers and $d=0.22\  10^{-3} m$ for polyamide. Finally, we set the experimental stresses as $\sigma_i=\frac{4F_i}{\pi d^2}$ and $E=920MPa$ or $E=2.5GPa$, for polyester and polyamide wires respectively. The parameters ML estimates were   $\mu=0.47 Gpa$, $\varepsilon_0 =0.0176$, $V0 =5.28\ 10^{-13} m^3$, $\omega_0=0.13 sec^{-1}$ in case of polyester, and  $\mu=1.06 Gpa$, $\varepsilon_0 =0.011$, $V_0 =1.68\ 10^{-13} m^3$, $\omega_0=0.002 sec^{-1}$ in case of polyamide fibers.

\begin{figure}[h]
\centering
\includegraphics[scale=.8]{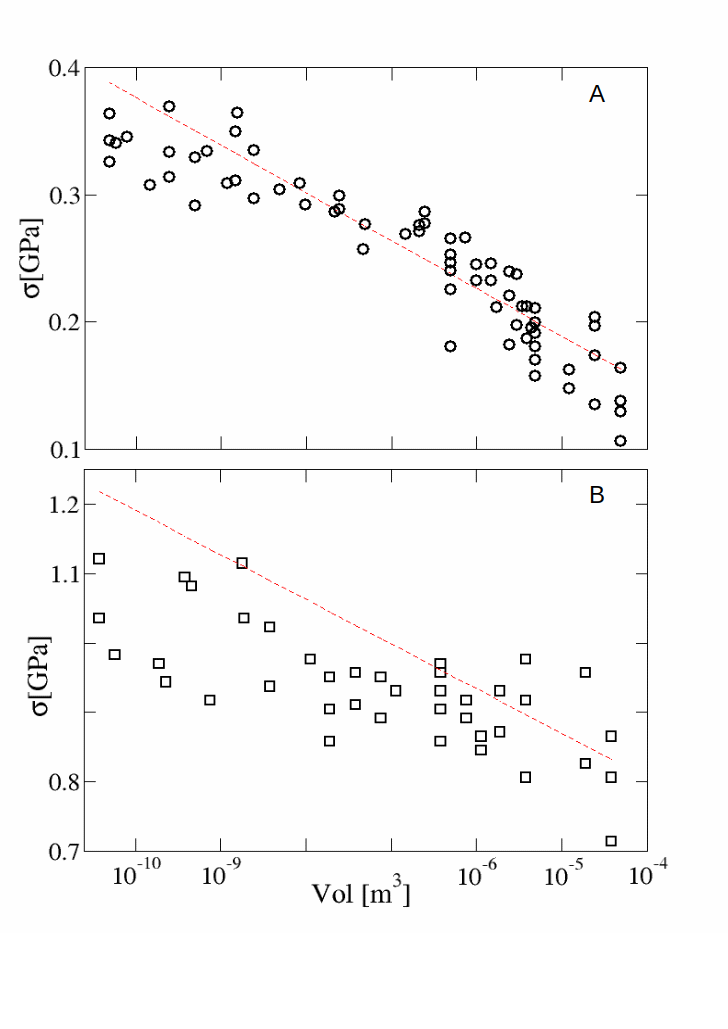}
\caption{Fracture stresses of polyester (A) and polyamide (B)  fibers from the experiments in \cite{fontana2020petersburg} (Fig.2). The ML parameters estimates were used to calculate $\langle \sigma\rangle$ (red dashed line).
}
\label{fig1}
\end{figure}

In Fig.\ref{fig1} the experimental fracture stresses (symbols) are plotted together with the average values $ \langle \sigma_{V_i}(\dot{\varepsilon_i};T)\rangle$(red dashed lines), defined as

\begin{equation}
\langle \sigma_{V_i}(\dot{\varepsilon_i};T)\rangle = \int_0^{\infty}d\sigma  \Sigma_{V_i}(\sigma;\dot{\varepsilon_i},T).
\label{avg_stress} 
 \end{equation}

\begin{figure}[h]
\centering
\includegraphics[scale=0.8]{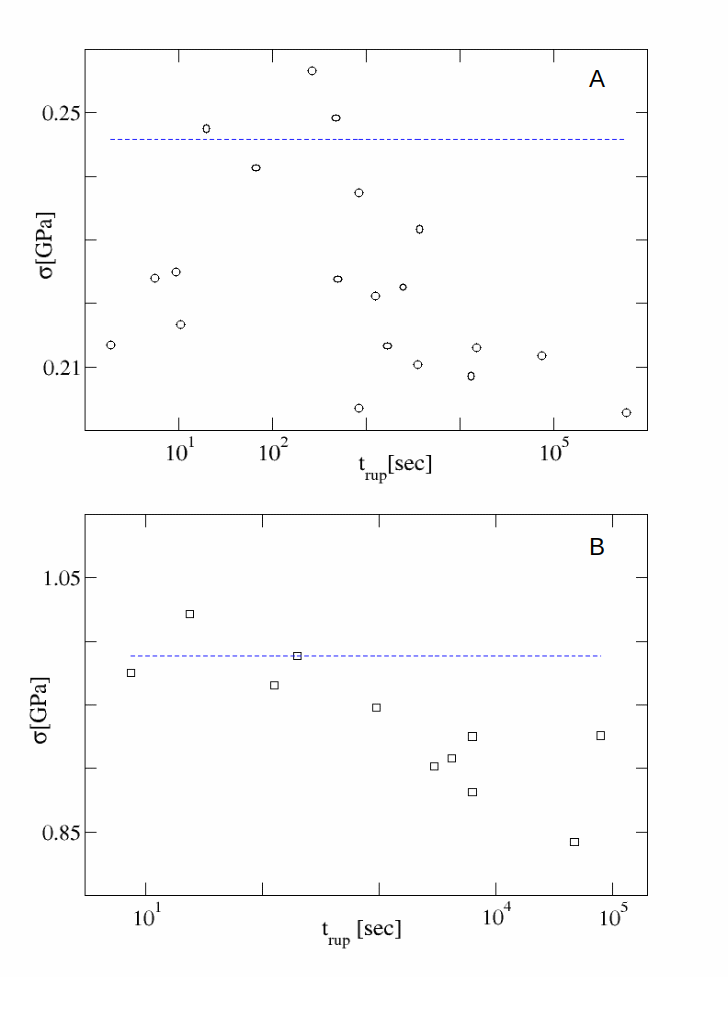}
\caption{Breaking stresses as a function of the rupture times  for polyester (A) and polyamide (B)  wires (Fig.3  in \cite{fontana2020petersburg}). Dashed blue lines are $\langle\sigma\rangle$, evaluated using the same parameters values fitted in the Fig.\ref{fig1}, and using $V=0.3\  10^{-6} m^3$ (polyester),  $V=0.14\  10^{-6}  m^3$ (polyamide) and $\dot{\varepsilon}_i =1/t^{i}_{rup}$. 
}
\label{fig2}
\end{figure}

The second set of experiments are reported in Fig.\ref{fig2} (symbols). We have plotted the average stress as a function of the rupture time $t_rup$, calculated from Eq.\ref{avg_stress} by adopting the parameters ML estimates found in the previous analysis. The wires' lengths were $L=6.2 m$ in case of polyester and $L=3.8 m$ for polyamides, while the strain rates could only be estimated indirectly by $1/t_{rup}^i$, as also  acknowledged by the authors \cite{fontana2020}.
Our theoretical prediction would indicate a very small strain dependence of the rupture forces in agreement with the experiments on polyester
(Fig.\ref{fig2} (A)). On the other side the experiments on polyamide indicate
that the exhibited strain dependence is not captured by the theoretical average stress. A possible reason for this discrepancy is that the precise value of the strain-rate is not known  and its estimate $1/t_{rup}$ is not correct.

\section{Conclusions}

We have shown how classical fracture mechanics, coupled to extreme value theory, provides an adequate theoretical framework to explain the results of tensile experiments performed on polyamide as well as on polyester wires reported in \cite{fontana2020petersburg}. We have shown that the Gumbel distribution reproduces the logarithmic decay of the average failure stress as a function of the wire length. We have used the generalized EVT to fit  the experimental failure stresses, showing no significant strain-rate dependence.

\bibliography{comment_Fontana}

\begin{thebibliography}{13}%
\makeatletter
\providecommand \@ifxundefined [1]{%
 \@ifx{#1\undefined}
}%
\providecommand \@ifnum [1]{%
 \ifnum #1\expandafter \@firstoftwo
 \else \expandafter \@secondoftwo
 \fi
}%
\providecommand \@ifx [1]{%
 \ifx #1\expandafter \@firstoftwo
 \else \expandafter \@secondoftwo
 \fi
}%
\providecommand \natexlab [1]{#1}%
\providecommand \enquote  [1]{``#1''}%
\providecommand \bibnamefont  [1]{#1}%
\providecommand \bibfnamefont [1]{#1}%
\providecommand \citenamefont [1]{#1}%
\providecommand \href@noop [0]{\@secondoftwo}%
\providecommand \href [0]{\begingroup \@sanitize@url \@href}%
\providecommand \@href[1]{\@@startlink{#1}\@@href}%
\providecommand \@@href[1]{\endgroup#1\@@endlink}%
\providecommand \@sanitize@url [0]{\catcode `\\12\catcode `\$12\catcode
  `\&12\catcode `\#12\catcode `\^12\catcode `\_12\catcode `\%12\relax}%
\providecommand \@@startlink[1]{}%
\providecommand \@@endlink[0]{}%
\providecommand \url  [0]{\begingroup\@sanitize@url \@url }%
\providecommand \@url [1]{\endgroup\@href {#1}{\urlprefix }}%
\providecommand \urlprefix  [0]{URL }%
\providecommand \Eprint [0]{\href }%
\providecommand \doibase [0]{https://doi.org/}%
\providecommand \selectlanguage [0]{\@gobble}%
\providecommand \bibinfo  [0]{\@secondoftwo}%
\providecommand \bibfield  [0]{\@secondoftwo}%
\providecommand \translation [1]{[#1]}%
\providecommand \BibitemOpen [0]{}%
\providecommand \bibitemStop [0]{}%
\providecommand \bibitemNoStop [0]{.\EOS\space}%
\providecommand \EOS [0]{\spacefactor3000\relax}%
\providecommand \BibitemShut  [1]{\csname bibitem#1\endcsname}%
\let\auto@bib@innerbib\@empty
\bibitem [{\citenamefont {Griffith}(1921)}]{griffith1921vi}%
  \BibitemOpen
  \bibfield  {author} {\bibinfo {author} {\bibfnamefont {A.~A.}\ \bibnamefont
  {Griffith}},\ }\bibfield  {title} {\bibinfo {title} {Vi. the phenomena of
  rupture and flow in solids},\ }\href@noop {} {\bibfield  {journal} {\bibinfo
  {journal} {Philosophical transactions of the royal society of london. Series
  A, containing papers of a mathematical or physical character}\ }\textbf
  {\bibinfo {volume} {221}},\ \bibinfo {pages} {163} (\bibinfo {year}
  {1921})}\BibitemShut {NoStop}%
\bibitem [{\citenamefont {Taloni}\ \emph {et~al.}(2018)\citenamefont {Taloni},
  \citenamefont {Vodret}, \citenamefont {Costantini},\ and\ \citenamefont
  {Zapperi}}]{taloni2018size}%
  \BibitemOpen
  \bibfield  {author} {\bibinfo {author} {\bibfnamefont {A.}~\bibnamefont
  {Taloni}}, \bibinfo {author} {\bibfnamefont {M.}~\bibnamefont {Vodret}},
  \bibinfo {author} {\bibfnamefont {G.}~\bibnamefont {Costantini}},\ and\
  \bibinfo {author} {\bibfnamefont {S.}~\bibnamefont {Zapperi}},\ }\bibfield
  {title} {\bibinfo {title} {Size effects on the fracture of microscale and
  nanoscale materials},\ }\href@noop {} {\bibfield  {journal} {\bibinfo
  {journal} {Nature Reviews Materials}\ }\textbf {\bibinfo {volume} {3}},\
  \bibinfo {pages} {211} (\bibinfo {year} {2018})}\BibitemShut {NoStop}%
\bibitem [{\citenamefont {Ball}(2020)}]{ball2020classic}%
  \BibitemOpen
  \bibfield  {author} {\bibinfo {author} {\bibfnamefont {P.}~\bibnamefont
  {Ball}},\ }\bibfield  {title} {\bibinfo {title} {Classic fail},\ }\href@noop
  {} {\bibfield  {journal} {\bibinfo  {journal} {Nature Materials}\ }\textbf
  {\bibinfo {volume} {19}},\ \bibinfo {pages} {829} (\bibinfo {year}
  {2020})}\BibitemShut {NoStop}%
\bibitem [{\citenamefont {Fontana}\ and\ \citenamefont
  {Palffy-Muhoray}(2020)}]{fontana2020petersburg}%
  \BibitemOpen
  \bibfield  {author} {\bibinfo {author} {\bibfnamefont {J.}~\bibnamefont
  {Fontana}}\ and\ \bibinfo {author} {\bibfnamefont {P.}~\bibnamefont
  {Palffy-Muhoray}},\ }\bibfield  {title} {\bibinfo {title} {St. petersburg
  paradox and failure probability},\ }\href
  {https://doi.org/10.1103/PhysRevLett.124.245501} {\bibfield  {journal}
  {\bibinfo  {journal} {Phys. Rev. Lett.}\ }\textbf {\bibinfo {volume} {124}},\
  \bibinfo {pages} {245501} (\bibinfo {year} {2020})}\BibitemShut {NoStop}%
\bibitem [{\citenamefont {Duxbury}\ \emph {et~al.}(1987)\citenamefont
  {Duxbury}, \citenamefont {Leath},\ and\ \citenamefont
  {Beale}}]{duxbury1987breakdown}%
  \BibitemOpen
  \bibfield  {author} {\bibinfo {author} {\bibfnamefont {P.}~\bibnamefont
  {Duxbury}}, \bibinfo {author} {\bibfnamefont {P.}~\bibnamefont {Leath}},\
  and\ \bibinfo {author} {\bibfnamefont {P.~D.}\ \bibnamefont {Beale}},\
  }\bibfield  {title} {\bibinfo {title} {Breakdown properties of quenched
  random systems: the random-fuse network},\ }\href@noop {} {\bibfield
  {journal} {\bibinfo  {journal} {Physical Review B}\ }\textbf {\bibinfo
  {volume} {36}},\ \bibinfo {pages} {367} (\bibinfo {year} {1987})}\BibitemShut
  {NoStop}%
\bibitem [{\citenamefont {Fisher}\ and\ \citenamefont
  {Tippett}(1928)}]{fisher1928limiting}%
  \BibitemOpen
  \bibfield  {author} {\bibinfo {author} {\bibfnamefont {R.~A.}\ \bibnamefont
  {Fisher}}\ and\ \bibinfo {author} {\bibfnamefont {L.~H.~C.}\ \bibnamefont
  {Tippett}},\ }\bibfield  {title} {\bibinfo {title} {Limiting forms of the
  frequency distribution of the largest or smallest member of a sample},\ }in\
  \href@noop {} {\emph {\bibinfo {booktitle} {Mathematical Proceedings of the
  Cambridge Philosophical Society}}},\ Vol.~\bibinfo {volume} {24}\ (\bibinfo
  {organization} {Cambridge University Press},\ \bibinfo {year} {1928})\ pp.\
  \bibinfo {pages} {180--190}\BibitemShut {NoStop}%
\bibitem [{\citenamefont {Gnedenko}(1943)}]{gnedenko1943distribution}%
  \BibitemOpen
  \bibfield  {author} {\bibinfo {author} {\bibfnamefont {B.}~\bibnamefont
  {Gnedenko}},\ }\bibfield  {title} {\bibinfo {title} {Sur la distribution
  limite du terme maximum d'une serie aleatoire},\ }\href@noop {} {\bibfield
  {journal} {\bibinfo  {journal} {Annals of mathematics}\ ,\ \bibinfo {pages}
  {423}} (\bibinfo {year} {1943})}\BibitemShut {NoStop}%
\bibitem [{\citenamefont {Leadbetter}\ \emph {et~al.}(2012)\citenamefont
  {Leadbetter}, \citenamefont {Lindgren},\ and\ \citenamefont
  {Rootz{\'e}n}}]{leadbetter2012extremes}%
  \BibitemOpen
  \bibfield  {author} {\bibinfo {author} {\bibfnamefont {M.~R.}\ \bibnamefont
  {Leadbetter}}, \bibinfo {author} {\bibfnamefont {G.}~\bibnamefont
  {Lindgren}},\ and\ \bibinfo {author} {\bibfnamefont {H.}~\bibnamefont
  {Rootz{\'e}n}},\ }\href@noop {} {\emph {\bibinfo {title} {Extremes and
  related properties of random sequences and processes}}}\ (\bibinfo
  {publisher} {Springer Science \& Business Media},\ \bibinfo {year}
  {2012})\BibitemShut {NoStop}%
\bibitem [{\citenamefont {Manzato}\ \emph {et~al.}(2012)\citenamefont
  {Manzato}, \citenamefont {Shekhawat}, \citenamefont {Nukala}, \citenamefont
  {Alava}, \citenamefont {Sethna},\ and\ \citenamefont
  {Zapperi}}]{manzato2012fracture}%
  \BibitemOpen
  \bibfield  {author} {\bibinfo {author} {\bibfnamefont {C.}~\bibnamefont
  {Manzato}}, \bibinfo {author} {\bibfnamefont {A.}~\bibnamefont {Shekhawat}},
  \bibinfo {author} {\bibfnamefont {P.~K.}\ \bibnamefont {Nukala}}, \bibinfo
  {author} {\bibfnamefont {M.~J.}\ \bibnamefont {Alava}}, \bibinfo {author}
  {\bibfnamefont {J.~P.}\ \bibnamefont {Sethna}},\ and\ \bibinfo {author}
  {\bibfnamefont {S.}~\bibnamefont {Zapperi}},\ }\bibfield  {title} {\bibinfo
  {title} {Fracture strength of disordered media: Universality, interactions,
  and tail asymptotics},\ }\href@noop {} {\bibfield  {journal} {\bibinfo
  {journal} {Physical review letters}\ }\textbf {\bibinfo {volume} {108}},\
  \bibinfo {pages} {065504} (\bibinfo {year} {2012})}\BibitemShut {NoStop}%
\bibitem [{\citenamefont {Kunz}\ and\ \citenamefont
  {Souillard}(1978)}]{kunz1978essential}%
  \BibitemOpen
  \bibfield  {author} {\bibinfo {author} {\bibfnamefont {H.}~\bibnamefont
  {Kunz}}\ and\ \bibinfo {author} {\bibfnamefont {B.}~\bibnamefont
  {Souillard}},\ }\bibfield  {title} {\bibinfo {title} {Essential singularity
  in the percolation model},\ }\href@noop {} {\bibfield  {journal} {\bibinfo
  {journal} {Physical Review Letters}\ }\textbf {\bibinfo {volume} {40}},\
  \bibinfo {pages} {133} (\bibinfo {year} {1978})}\BibitemShut {NoStop}%
\bibitem [{\citenamefont {Gumbel}(2004)}]{gumbel2004statistics}%
  \BibitemOpen
  \bibfield  {author} {\bibinfo {author} {\bibfnamefont {E.~J.}\ \bibnamefont
  {Gumbel}},\ }\href@noop {} {\emph {\bibinfo {title} {Statistics of
  extremes}}}\ (\bibinfo  {publisher} {Courier Corporation},\ \bibinfo {year}
  {2004})\BibitemShut {NoStop}%
\bibitem [{\citenamefont {Sellerio}\ \emph {et~al.}(2015)\citenamefont
  {Sellerio}, \citenamefont {Taloni},\ and\ \citenamefont
  {Zapperi}}]{sellerio2015fracture}%
  \BibitemOpen
  \bibfield  {author} {\bibinfo {author} {\bibfnamefont {A.~L.}\ \bibnamefont
  {Sellerio}}, \bibinfo {author} {\bibfnamefont {A.}~\bibnamefont {Taloni}},\
  and\ \bibinfo {author} {\bibfnamefont {S.}~\bibnamefont {Zapperi}},\
  }\bibfield  {title} {\bibinfo {title} {Fracture size effects in nanoscale
  materials: the case of graphene},\ }\href@noop {} {\bibfield  {journal}
  {\bibinfo  {journal} {Physical Review Applied}\ }\textbf {\bibinfo {volume}
  {4}},\ \bibinfo {pages} {024011} (\bibinfo {year} {2015})}\BibitemShut
  {NoStop}%
\bibitem [{\citenamefont {Fontana}()}]{fontana2020}%
  \BibitemOpen
  \bibfield  {author} {\bibinfo {author} {\bibfnamefont {J.}~\bibnamefont
  {Fontana}},\ }\href@noop {} {}\bibinfo {howpublished} {personal
  communication}\BibitemShut {NoStop}%
\end{thebibliography}%

\end{document}